\pdfoutput=1
\documentclass[]{spie}  

\usepackage{amsmath,amsfonts,amssymb}
\usepackage{graphicx}
\usepackage[colorlinks=true, allcolors=blue]{hyperref}

\title{Pulse processing in TES detectors: comparison of different short filter methods based on optimal filtering. Case study for Athena X-IFU}

\author[a]{Beatriz Cobo}
\author[b]{Nicol\'as Cardiel}
\author[a]{Mar\'{\i}a Teresa Ceballos}
\author[c]{Philippe Peille}

\affil[a]{Instituto de F\'{\i}sica de Cantabria (CSIC-UC), Edificio Juan Jord\'a, Avenida de los Castros, s/n - E-39005 Santander, Cantabria, Spain}
\affil[b]{Departamento de F\'{\i}sica de la Tierra y Astrof\'{\i}sica, Universidad Complutense de Madrid, Plaza Ciencias 1, E-28040 Madrid, Spain}
\affil[c]{CNES, 18 Av. Edouard Belin, 31401 Toulouse Dedex 9, France}

\authorinfo{Further author information: (Send correspondence to Beatriz Cobo)\\Beatriz Cobo: E-mail: cobo@ifca.unican.es}

\begin{document} 
\maketitle

\begin{abstract}
In the framework of the ESA Athena mission, the X-ray Integral Field Unit (X-IFU) instrument to be on board the X-ray Athena Observatory is a cryogenic micro-calorimeter array of Transition Edge Sensor (TES) detectors aimed at providing spatially resolved high-resolution spectroscopy. As a part of the on-board Event Processor (EP), the reconstruction software will provide the energy, spatial location and arrival time of the incoming X-ray photons hitting the detector and inducing current pulses on it. Being the standard optimal filtering technique the chosen baseline reconstruction algorithm, different modifications have been analyzed to process pulses shorter than those considered of high resolution (those where the full length is not available due to a close pulse after them) in order to select the best option based on energy resolution and computing performance results. It can be concluded that the best approach to optimize the energy resolution for short filters is the 0-padding filtering technique, benefiting also from a reduction in the computational resources. However, its high sensitivity to offset fluctuations currently prevents its use as the baseline treatment for the X-IFU application for lack of consolidated information on the actual stability it will get in flight.

\end{abstract}

\keywords{Athena, X-ray Integral Field Unit, X-ray spectroscopy, space telescopes, instrumentation, reconstruction methods, performance analysis}

\section{INTRODUCTION}
\label{sec:intro}  

The X-ray Integral Field Unit (X-IFU)\cite{Barret2018} instrument that will be on-board the ESA's Athena mission\cite{Nandra2013}, is a high-resolution cryogenic imaging spectrometer in the 0.2-12 keV band that will provide unprecedented spectral resolution (2.5 eV at 7 keV) with 5" of spatial resolution. The X-IFU Focal Plane will contain a large array of Transition Edge Sensors (TES) (3168 pixels) in groups of several tens of TES per readout channel considering a Time Division Multiplexing (TDM) scheme. The on-board Event Processor (EP)\cite{Ravera2014} is separated in two parts: an FPGA-based trigger will perform the triggering of the X-ray pulses and after this, a scientific software implemented in a space qualified processor will estimate the energy and the arrival time of the detected events providing also the spatial location (based on impact pixel).

The main objective of the pulse processing is to get the best energy resolution from the photons arriving at a X-IFU TES detector. Therefore, the selected algorithms to work with them should be optimal and capable of reducing the degradation of the energy resolution with energy due to the non-linearity of the detector. 

On one hand, a performance assessment of two different triggering algorithms\cite{Cobo2018} led to the selection of the Single Threshold Crossing technique as the current detection baseline. On the other hand, a similar comparative analysis of several pulse reconstruction algorithms\cite{Peille2016} established the optimal filtering technique {\cite{Moseley1988,Szymkowiak1993,Boyce1999}} applied to a transformed quasi-resistance space\cite{Bandler2006} as the current reconstruction baseline. In both cases (the triggering and the reconstruction) the best compromise between the detection efficiency and the energy resolution and computational and calibration costs was the criteria used to the selection of the optimal methods.

In this paper we present the optimization of the standard optimal filtering technique in the case of short filters, those used when photons hitting the detector are arriving so close that choosing the longest, high resolution filter is not possible. Therefore, the performance of two new approaches to the reconstruction algorithms will be assessed in terms of the energy resolution achieved, the computing power required and the stability under instrumental offset fluctuations.

First, a revision of the reconstruction baseline will be briefly shown presenting some potential improvements (in Sec.~\ref{sec:recons_tech}). These methods are applied to synthetic data streams simulated with the XIFUSIM\cite{Lorenz2020} simulator referring to an LPA2.5a pixel\cite{Ullom2018}, so a description of the simulation environment is done in Sec.~\ref{sec:simulation_frame}. Then, Sec.~\ref{sec:results} includes the results obtained in terms of energy resolution for the proposed modifications to the standard optimal filtering algorithm and finally, in Sec.~\ref{sec:sensitivity} the response to instrumental offset fluctuations is analysed.

Along this paper the following magnitudes will be used:
\begin{itemize}
\item{Sampling rate: 156250 kHz.}
\item{Reconstructed PH: pulse height returned by the reconstruction method (a proxy for the pulse energy).}
\item{Calibrated energy: energy of the pulse obtained after the gain scale correction is applied to the reconstructed PH.}
\item{Energy resolution:}
\end{itemize}

\begin{equation}
\label{eq:fwhm}
FWHM = 2.35 \cdot \sqrt{\frac{(E_i-{\bar{E}})^2}{N-1}} \, ,
\end{equation}
where $E_i$ is the calibrated energy of a reconstructed pulse and $N$ is the number of pulses at a given energy, used for the resolution calculation.

\begin{itemize}
\item{Grading: pulses are graded according to the record length available for their reconstruction and thus to the energy resolution that could provide\cite{Peille2018}.}
\end{itemize}

\begin{table}[ht]
\caption{Triggered pulses shall be graded according to this table.} 
\label{tab:grading}
\begin{center}
\begin{tabular}{|c|c|c|c|} 
\hline
\rule[-1ex]{0pt}{3.5ex}  \textbf{GRADE} & \textbf{Time until} & \textbf{Time since} & \textbf{Energy resolution} \\
\rule[-1ex]{0pt}{3.5ex}  \textbf{} & \textbf{next pulse} & \textbf{previous pulse} & \textbf{@ 7 keV (FWHM)} \\
\hline
\rule[-1ex]{0pt}{3.5ex}  { Very High Resolution, VHR} & {$>$8192 samples} & {$>$494 samples} & {2.5 eV}   \\
\rule[-1ex]{0pt}{3.5ex}  {High Resolution, HR} & {$>$4096 samples} & {$>$494 samples} & {$>$$\sim$2.5 eV}   \\
\rule[-1ex]{0pt}{3.5ex}  {Intermediate Resolution, IR} & {$>$2048 samples} & {$>$494 samples} & {$\sim$2.6 eV}   \\
\rule[-1ex]{0pt}{3.5ex}  {Medium Resolution, MR} & {$>$512 samples} & {$>$494 samples} & {$\sim$3 eV}   \\
\rule[-1ex]{0pt}{3.5ex}  {Limited Resolution} & {$>$256 samples} & {$>$494 samples} & {$\sim$7 eV}   \\
\rule[-1ex]{0pt}{3.5ex}  {Low Resolution} & {$>$8 samples} & {$>$494 samples} & {$\sim$30 eV}   \\
\hline
\end{tabular}
\end{center}
\end{table}

\section{RECONSTRUCTION TECHNIQUES}
\label{sec:recons_tech}

\subsection{Optimal Filtering} 
\label{sec:opt_filter}

The Optimal filter is the standard reconstruction technique designed to get the best estimate of the energy of a photon in a microcalorimeter. It relies on two main assumptions:

Firstly, the detector response is linear; that is, the pulse shapes are identical regardless of their energy and thus, the pulse amplitude is the scaling factor from one pulse to another\cite{Szymkowiak1993}. In the frequency domain (as noise can be frequency dependent), the raw data can be expressed as

\begin{equation}
\label{eq:d(f)}
D(f) = E \cdot S(f) + N(f) \, 
\end{equation}
where $S(f)$ is the normalized model pulse shape (matched filter), $N(f)$ is the noise and $E$ is the scalar amplitude for the photon energy.

The second assumption is that the noise is stationary, i.e. it does not vary with time. The amplitude of each pulse can then be estimated by minimizing (weighted least-squares sense) the difference between the noisy data and the model pulse shape, being the $\chi$² condition to be minimized:

\begin{equation}
\label{eq:chi2}
\chi^2=\int_{-\infty}^{\infty}\frac{(D(f)-E \cdot S(f))^2}{<(N(f)^2)>}df \, 
\end{equation}
	
In the time domain, the amplitude is the best weighted (optimally filtered) sum of the values in the pulse

\begin{equation}
\label{eq:E}
E=k\int_{-\infty}^{\infty}D(t) \cdot OF(t)dt \, 
\end{equation}
where $OF(t)$ is the time domain expression of optimal filter in frequency domain

\begin{equation}
\label{eq:OF(f)}
OF(f)=\frac{S^{*}{(f)}}{<(N(f)^2)>} \, 
\end{equation}
and $k$ is the normalization factor to give $E$ in units of energy

\begin{equation}
\label{eq:k}
k=\int_{-\infty}^{\infty} \frac{S{(f)} \cdot S^{*}{(f)}}{<(N(f)^2)>}df \, 
\end{equation}
	
For this method, noise is typically characterized by its power spectral density (PSD) and then the filter can be constructed using the Discrete Fourier Transform (DFT).

\subsection{Short Filters Optimal Filtering} 
\label{sec:short_filters}

Optimal filters are built in principle from the largest pulse shape available, i.e. the length considered for the Very High resolution pulses. However, the astronomical photons reaching the detector can arrive at different separations from each other, giving pulses of different (sub-optimal) processing lengths. 

The usual approach to the processing of such shorter pulses is by classifying all the pulses in a small number of grades (see Table.~\ref{tab:grading}) and reconstructing them using pre-calculated shorter filters.

A filter is built (see Eq.~\ref{eq:OF(f)}) using a pulse template and a description of the noise, so the usual approach to create a short filter is thus, using templates of the required lengths (8192, 4096, 2048, 512, 256 and 8 for the presented grading scheme). However, when these short filters are used to reconstruct small lengths the energy resolution of the reconstruction achieves undesirable high values. 

\begin{figure}[ht]
\begin{tabular}{cc}
\includegraphics[scale=0.24]{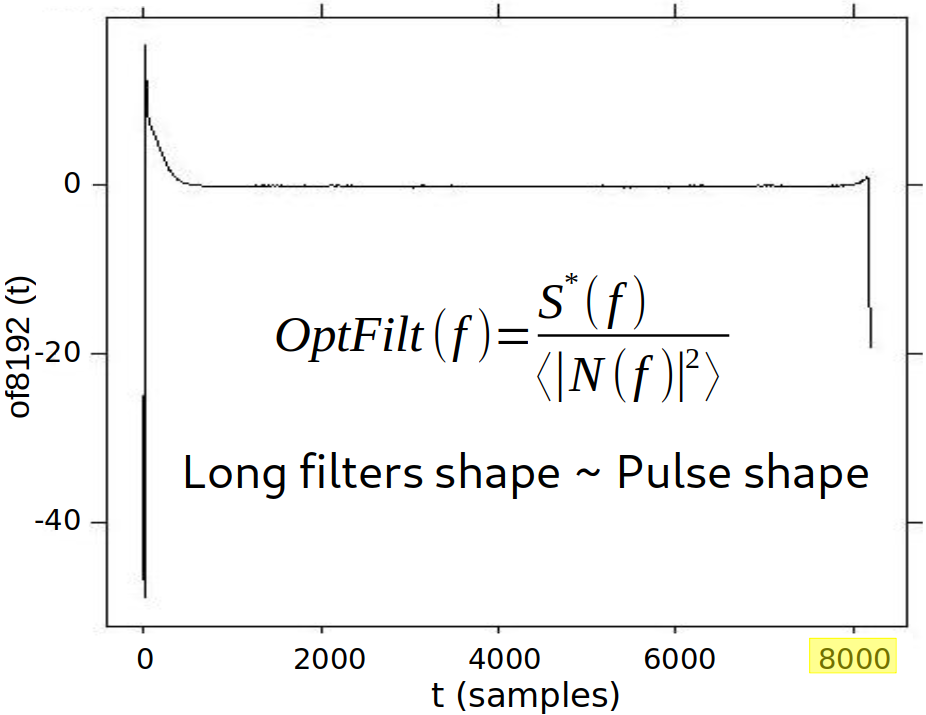}
&
\includegraphics[scale=0.24]{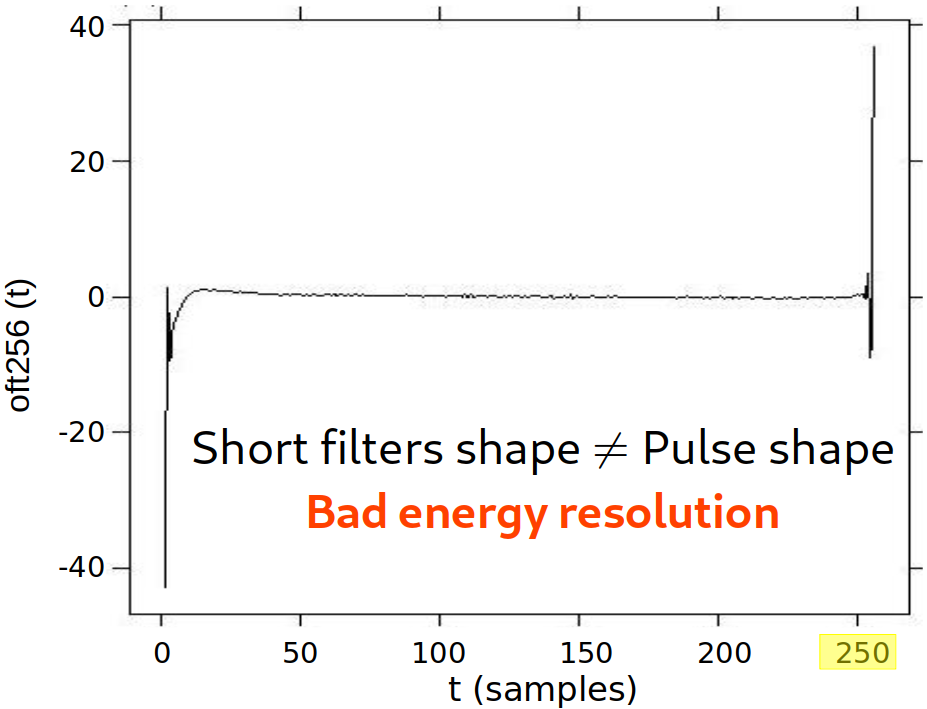}
\end{tabular}
\caption{Left: Long optimal filter in Time Domain. 
Right: Short optimal filter in Time Domain. 8192 samples is the Very High Resolution filter length and 256 samples is the Limited Resolution filter length (8000 and 250 are highlighted in yellow to pay attention to the scale of both plots).}
\label{fig:shortFilters}
\end{figure}

In Fig.~\ref{fig:shortFilters} both long and short filters are plot in the Time Domain. According again to Eq.~\ref{eq:OF(f)} and assuming white noise, the shape of the filter should match the shape of the pulse whose energy is going to be reconstructed (plot on the left) which is not true in the case of short filters (plot on the right) causing the energy resolution degradation.   

\begin{figure}[h]
\centering
\begin{tabular}{cc}
\includegraphics[scale=0.15]{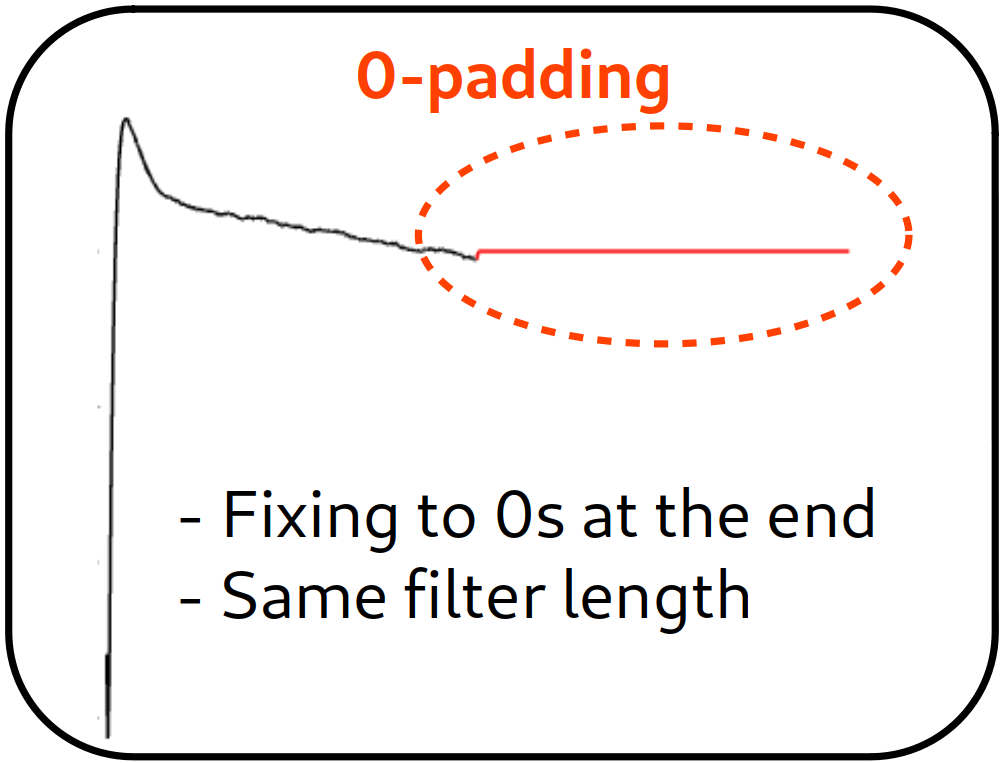}
&
\includegraphics[scale=0.15]{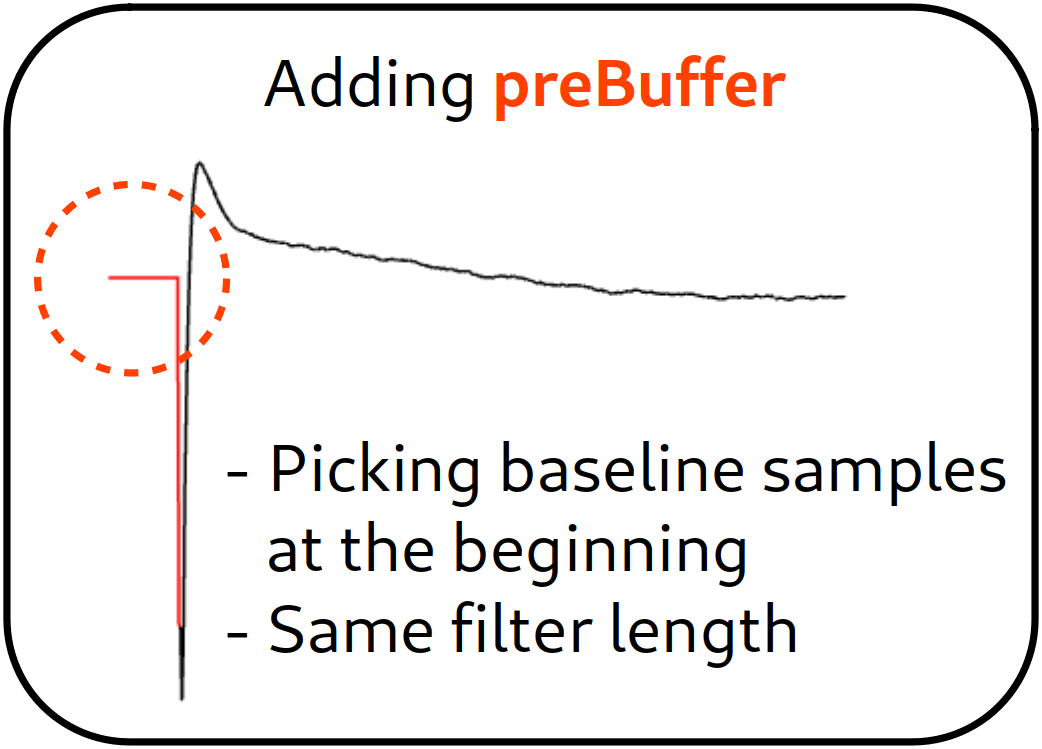}
\end{tabular}
\caption{Left: 0-padding. 
Right: Pre-buffer addition.}
\label{fig:alternatives}
\end{figure}

To avoid this behavior, a different approach known as \textbf{0-padding} can be used. Instead of using a reduced-length template to build the different short filters, the initial filter built using a very high resolution-long template is always used (left plot of Fig.~\ref{fig:alternatives}). For shorter ($s$ samples) pulses, only the initial $s$ samples of the Time Domain filter are used in the Time Domain scalar product:
  
\begin{equation}
\label{eq:0padding}
E \sim \sum_{i=1}^{s} d(t)_i \cdot OF(t)_i \, 
\end{equation} 
where the upper limit $s$ is the length of the pulse

An alternative also explored in this paper is the \textbf{addition of a pre-buffer} signal to the pulse template before building the filter, i.e. using some extra samples before the triggering point of the event (right plot of Fig.~\ref{fig:alternatives}). When the reconstruction is performed, the detected event is also stored with the same number of pre-buffer samples and the scalar product is done with the total number of samples assigned by the grading table (a Medium Resolution reconstruction will be performed with a pulse and a filter with $x$ pre-buffer samples + $(512-x)$ event/template samples so that the final length is 512 samples). 

A further modification of the optimal filtering achieved with the addition of a pre-buffer to the 0-padding approach is also explored.

\subsection{Quasi Resistance Space} 
\label{sec:resistance_space}

In some previous works\cite{Peille2016,Bandler2006}, a transformation of the input signal from the current space to a proxy for the resistance signal has been proposed to gain a more linear scaling of the optimal filtering reconstruction with the energy. 

Here, the most recent version of the "fitted" resistance transformation\cite{Peille2016} is:

\begin{equation}
\label{eq:Ifitted}
\frac{R}{V0} = \frac{1}{I_{fit}+I} \, 
\end{equation} 
where $I$ is used in arbitrary units, a.u. (values directly read from the simulated data current). $I_{fit}$ is the optimized value to get the closest-to-linear gain scale.

\section{SIMULATION FRAMEWORK}
\label{sec:simulation_frame}
\textit{}
This study is based on the simulations performed with the XIFUSIM\cite{Lorenz2020} simulator, a tool to provide a full model of the detection chain of the X-IFU instrument on \textit{Athena}, being a faithful and validated representation of the detection chain from the detection of photons in the TES pixels, through the analogue processing in the readout chain, up to the trigger and post processing in the instrument's digital electronics. In Tab.~\ref{tab:xifusim} the XIFUSIM instrument parameters used for the different studies presented in this paper are specified.

\begin{table}[ht]
\caption{XIFUSIM parameters used.} 
\label{tab:xifusim}
\begin{center}
\begin{tabular}{|l|c|} 
\hline
\rule[-1ex]{0pt}{3.5ex}  \textbf{XIFUSIM parameter} & \textbf{Value} \\
\hline
\rule[-1ex]{0pt}{3.5ex}  {Squid noise} & {9.62 pA/sqrt(Hz)}  \\
\rule[-1ex]{0pt}{3.5ex}  {Bias noise} & {0.3077 pA/sqrt(Hz)}  \\
\rule[-1ex]{0pt}{3.5ex}  {XIFUSIM version} & {0.7.1}  \\
\rule[-1ex]{0pt}{3.5ex}  {XIFUSIM instruments version} & {0.7.0} \\
\hline
\end{tabular}
\end{center}
\end{table}

The data used in this analysis have been simulated using the pixel parameters of Tab.~\ref{tab:pixel} in order to get representative X-IFU pixels, with a flux-locked loop (FLL) circuit (see Tab.~\ref{tab:fll}). 

\begin{table}[ht]
\hspace{0.5em}
\parbox{0.4\linewidth}{
\caption{Pixel parameters (LPA2.5a pixel).}
\label{tab:pixel}
\begin{center}
\begin{tabular}{|l|c|} 
\hline
\rule[-1ex]{0pt}{3.5ex}  \textbf{Pixel parameter} & \textbf{Value} \\
\hline
\rule[-1ex]{0pt}{3.5ex}  {Heat capacity at bias, C} & {0.726 pJ/K}  \\
\rule[-1ex]{0pt}{3.5ex}  {Bath conductance at bias, G} & {72.44 pW/K}  \\
\rule[-1ex]{0pt}{3.5ex}  {Heat bath power flow exponent, n} & {3.377}  \\
\rule[-1ex]{0pt}{3.5ex}  {$\alpha$} & {619.14}  \\
\rule[-1ex]{0pt}{3.5ex}  {$\beta$} & {21.9}  \\
\rule[-1ex]{0pt}{3.5ex}  {Bias resistance, $R_0$} & {0.9715 m$\Omega$}  \\
\rule[-1ex]{0pt}{3.5ex}  {Transition temperature, $T_0$} & {89.42 mK}  \\
\rule[-1ex]{0pt}{3.5ex}  {Bias current, $I_0$} & {39.9 $\mu$A}  \\
\rule[-1ex]{0pt}{3.5ex}  {Effective load resistance, $R_L$} & {75 $\mu\Omega$}  \\
\rule[-1ex]{0pt}{3.5ex}  {Effective circuit inductance, L} & {640 nH}  \\
\rule[-1ex]{0pt}{3.5ex}  {Transformer Turns Ratio, TTR} & {1}  \\
\hline
\end{tabular}
\end{center}
}
\hspace{7em}
\parbox{.4\linewidth}{
\caption{FLL parameters used.}
\label{tab:fll}
\begin{center}
\begin{tabular}{|l|c|} 
\hline
\rule[-1ex]{0pt}{3.5ex}  \textbf{FLL parameter} & \textbf{Value} \\
\hline
\rule[-1ex]{0pt}{3.5ex}  {DELAY} & {1 clock cycle} \\
\rule[-1ex]{0pt}{3.5ex}  {BIASLEAK} & {0} \\
\rule[-1ex]{0pt}{3.5ex}  {FB$\_$LEAK} & {0} \\
\rule[-1ex]{0pt}{3.5ex}  {FB$\_$LAG} & {1clock cycle} \\
\rule[-1ex]{0pt}{3.5ex}  {K$\_$I} & {0.2947} \\
\rule[-1ex]{0pt}{3.5ex}  {K$\_$MIX} & {0.1754} \\
\rule[-1ex]{0pt}{3.5ex}  {HIERARCH DAC$\_$WIDTH} & {14} \\
\rule[-1ex]{0pt}{3.5ex}  {DAC$\_$OMIN} & {0} \\
\rule[-1ex]{0pt}{3.5ex}  {DAC$\_$OMAX} & {0.0005} \\
\rule[-1ex]{0pt}{3.5ex}  {FB$\_$START} & {9299 a.u.} \\
\rule[-1ex]{0pt}{3.5ex}  {V$\_$LOCK} & {2048 a.u.} \\
\hline
\end{tabular}
\end{center}
}
\end{table}

\subsection{Precalculated Libraries} 
\label{sec:libraries}

A library is an external file containing a collection of precalculated values that would be used by the reconstruction algorithm (pulse templates, optimal filters, etc). 

The testing of the performance of the different reconstruction algorithms requires the creation of the appropriate library taking also into account the signal space (current or \textit{quasi}-resistance). For all the analyses, the energy for the filters is 6~keV. Such a library contains:

\begin{itemize}
\item{A pulse template built averaging 20000 very high resolution, perfectly isolated (closest neighbour more than 250 ms away) 6~keV pulses simulated at random offsets with the sampling process between $-0.5$ and $+0.5$ (photons have been placed at a random phase with the sampling rate -jitter- that is present in the real data).}
\item{The corresponding optimal filters at power of 2 lengths, from Very High Resolution (8192 samples) to 4 samples. They are built from the pulse template and the noise power spectrum, which is obtained by averaging the Discrete Fourier Transform of 1000 pulse-free records of length 8192 samples.}
\end{itemize}

It needs to be noted that, in case of using the addition of a particular pre-buffer to either the optimal filtering or the 0-padding methods, a new library taking into account the given pre-buffer should be created.

\subsection{Gain Scale} 
\label{sec:gainscale}

The assumption of single pulse shape for all energies that is made in optimal filtering reconstruction techniques is not verified in any large band TES detectors (non linear). Because of this, the result of the reconstruction  (pulse height) must be transformed into an unbiased estimation of the energy, in proper energy units, applying a gain scale (functional relation between the input simulated energy and the output reconstructed pulse height).

The most simple approach is a one-dimensional polynomial curve plotting the reconstructed pulse heights versus the input simulated energies and fitting by using a deg-6 polynomial. For this purpose 5000 high resolution perfectly isolated pulses have been simulated at calibration energies of 0.2, 0.5 1, 2, 3, 4, 5, 6, 7, 8, 9, 10, 11 and 12 keV and with random offsets between -0.5 and 0.5, and then filtered with a 6 keV optimal filter. This functional relationship must be derived for every reconstruction method and every filter length. Calibration is applied as E(keV) = f(PH). 
                       
\begin{figure} [ht]
\begin{center}
\includegraphics[width=12.5cm]{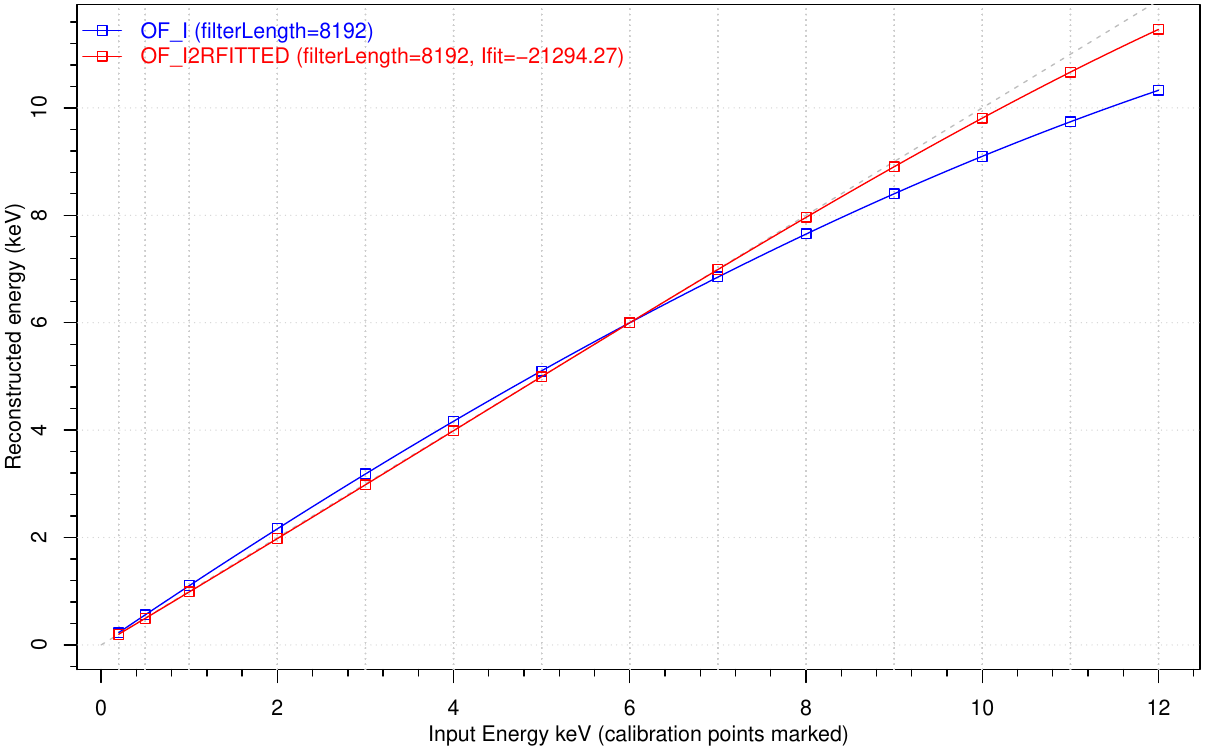}
\end{center}
\caption{Points: reconstructed pulse heights for the input calibration energies. Optimal filter at 6 keV in current space (blue) and in fitted resistance space (red). Lines: polynomial fits to the reconstructed points.}
\label{fig:gainScale} 
\end{figure}

The gain scales corresponding to optimal filtering in the current space and the proposed transformation to the quasi resistance space (shown in Sec.~\ref{sec:opt_filter} and ~\ref{sec:resistance_space}) are plot in Fig.~\ref{fig:gainScale}.

%
\subsection{Test Pulses} 
\label{sec:test_pulses}

The evaluation of the different reconstruction mechanisms is performed over 5000 very high resolution perfectly isolated simulated pulses of 7 keV with random offsets between -0.5 and 0.5. They are used to show the energy resolution achieved (FWHM) after an initial parabola correction (reconstructed pulse height is the maximum of the parabola given by the reconstructions performed at three different samples around the trigger time\cite{Peille2016}) and the gain scale calibration are applied. 

\section{RESULTS}
\label{sec:results}

\subsection{Selection of Prebuffer} 
\label{sec:prebuffer}

\begin{figure} [ht]
\begin{center}
\includegraphics[height=8cm]{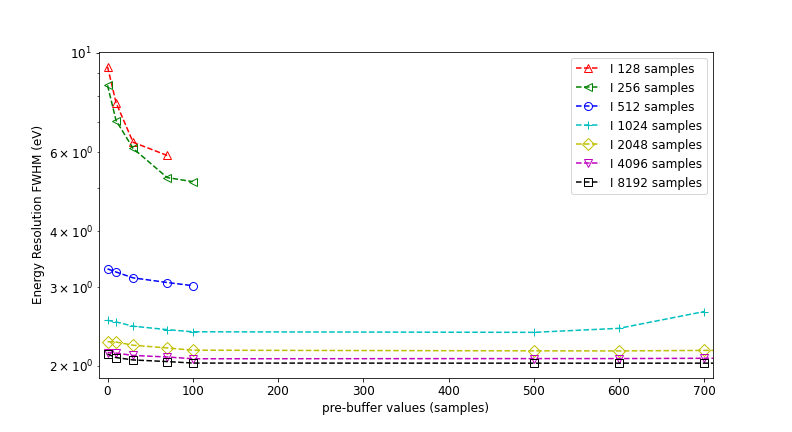}
\end{center}
\caption{Energy resolution for pre-buffer optimal filtering in current space (I). The length of the filters runs from 128 (red triangles) to 8192 (black squares) samples. The pre-buffer values tested are marked by the presence of symbols.} 
\label{fig:preBuffersAnalysis} 
\end{figure}

The pre-buffer approach to deal with short filters requires the selection of the optimal pre-buffer length value.  Fig.~\ref{fig:preBuffersAnalysis} shows the comparison of the energy resolution ($Y$ axis) obtained with different pre-buffer values ($X$ axis) at each filter length (see legend). The reconstruction has been performed in the current space with pulses of 7~keV. Only pre-buffer lengths shorter than 100 samples have been considered for the shortest filters.

As it can be seen, different values of the pre-buffer are optimal for different filter lengths and thus, these optimal values are the ones considered in the global comparison. For instance, 100 samples is the best pre-buffer for a 512-length filter while for a 1024-length filter the best pre-buffer is 500 samples.

\subsection{Comparison} 
\label{sec:comparison}

\begin{figure} [ht]
\begin{center}
\includegraphics[height=8cm]{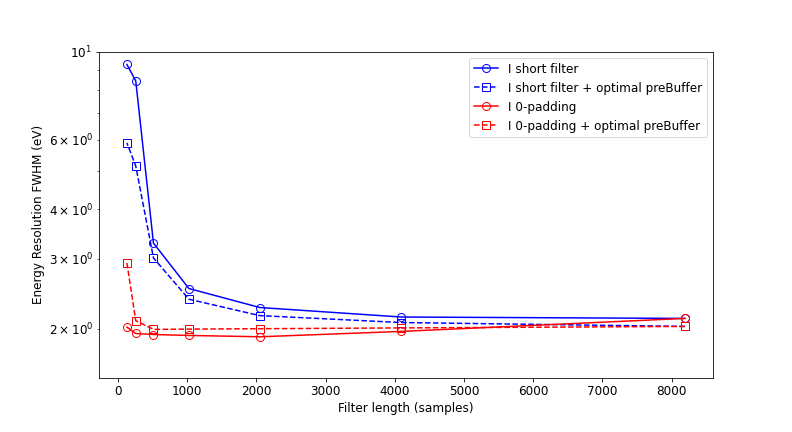}
\end{center}
\caption{Energy resolution performance of the different pulse reconstruction methods over 5000 test LPA2.5a calibrated pulses. An optimal pre-buffer for each filter length has been used (for example 70 samples of pre-buffer for a 128-samples-length filter but 600 samples of pre-buffer for a 8192-samples-length filter).} 
\label{fig:fwhm} 
\end{figure}

The performance of the different techniques is shown in Fig.~\ref{fig:fwhm}. The comparison includes short filters and 0-padding filters, with and without a pre-buffer, applied to the test pulses described in Sec. \ref{sec:test_pulses}.

Despite the pre-buffer approach with short filters meeting the energy resolution requirements for all the resolution pulses, the 0-padding approach (with/out pre-buffer) gives the best global resolution results with the additional bonus of a reduced computational power. According to these results, at 7 keV even Limited Resolution pulses (those with record lengths larger than 256 samples and shorter than 512 samples) could be reconstructed giving an energy resolution consistent with that of the Very High Resolution ones. The addition of a pre-buffer to the 0-padding filters does not improve the resolution results.

\section{SENSITIVITY TO INSTRUMENTAL VARIATIONS}
\label{sec:sensitivity}

In the standard optimal filtering technique (full length filter) the 0Hz bin of the Fourier Transform is not used to avoid the presence of 1/f noise caused by fluctuations generated in the readout electronics. That way, the final filter sum is zero and the baseline is suppressed from the processing. 

However, in the 0-padding approach where the filter is cut (in the Time Domain), the filter sum is not zero anymore and thus, the method needs to be tested against possible variations in the instrument parameters that could lead to an unaffordable energy resolution degradation or that otherwise would impose very restrictive conditions to the instrumental setup.

Amongst the possible changing conditions, sensitivity to the overall baseline level and to the fluctuations of the bath temperature, the baseline or the bias voltage should be studied.

\subsection{Baseline Level} 
\label{sec:baseline}

\begin{figure} [ht]
\begin{center}
\includegraphics[height=8cm]{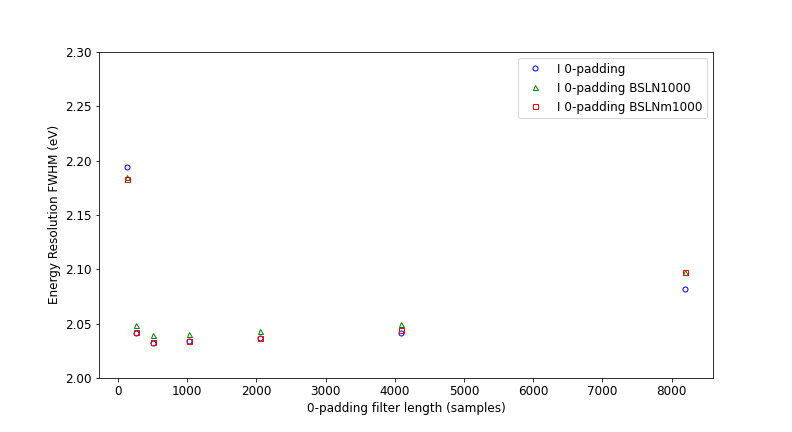}
\end{center}
\caption{Energy resolution (FWHM) of pulses of 7 keV as a function of the length of 0-padding filters. Baseline values are $\sim$6000 a.u. (BSLNm1000), $\sim$7000 a.u. (default) and $\sim$8000 a.u. (BSLN1000).} 
\label{fig:baseline} 
\end{figure}

The instrumental baseline level is a setup choice but it could be at levels where the 0-padding method could be affected. To check the severity of this effect, the baseline of our previous simulated test pulses has been lowered to $\sim$6000 a.u. and risen to $\sim$8000 a.u.(values for the simulator framework in \ref{tab:xifusim} are $\sim$7000 a.u.) and then the 0-padding reconstruction and posterior calibration have been performed again. Fig.~\ref{fig:baseline} shows the energy resolution obtained in both cases in comparison with the results for the initial (default baseline) simulations: values are independent of the baseline level selected.

\subsection{Bias Voltage Fluctuations} 
\label{sec:biasvoltage}

The behaviour of the 0-padding technique is checked for changes in the bias voltage level. For this purpose, two noise-and-jitter-free pulses of 7 keV have been simulated with XIFUSIM, one for the default bias voltage ($V0=4.175535 \times 10^{-8}$~V) and the other one for an V0 1ppm larger. In addition, noise-and-jitter-free pulses at the calibration energies and at the default bias voltage have been also simulated to fit the gain scale curve. These pulses have been then reconstructed with the standard optimal filtering technique and using 0-padding at different filter lengths.

\begin{figure} [ht]
\begin{center}
\includegraphics[height=7.5cm]{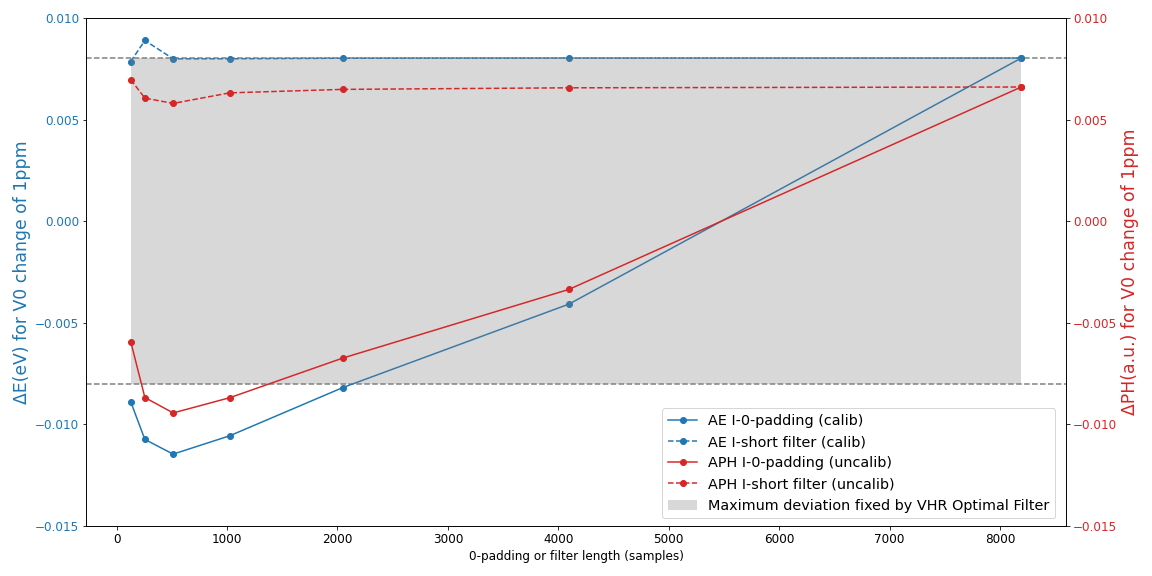}
\end{center}
\caption{0-padding (continuous lines) and short filter (dashed lines) sensitivity to a 1ppm change in the bias voltage. Blue lines represent a change in calibrated energies (left axis, $\Delta$E) and red lines changes in uncalibrated pulse height values (right axis, $\Delta$PH). Shaded area dimensions the region for maximum deviation given by the standard (full length) optimal filtering technique.} 
\label{fig:biasvoltage} 
\end{figure}

Fig.~\ref{fig:biasvoltage} shows the comparison of the variation in the reconstructed PH (before calibration; right red axis) and in the reconstructed energy (after calibration; left blue axis) for both approaches. The gain scale polynomial to perform the calibration has been fitted to the reconstructed Pulse Height of 1 single pulse of every energy (simulated without noise and without jitter).

According to these results, the 0-padding approach with records lengths of 4096 and even 2048 samples would have a sensitivity to bias voltage fluctuations comparable to those shown by the standard (full length 8192 samples) optimal filtering, thus keeping the performance advantages stated in Sec.~\ref{sec:comparison}.

\subsection{Bath-Temperature-Induced Baseline Fluctuations} 
\label{sec:tbath}

Following the study of the bias voltage sensitivity, a change in the bath temperature can induce a variability in the baseline level that affects the reconstruction process. To get the magnitude of this change, two pulses have been simulated with XIFUSIM, one at the base temperature of 55mK and the other one a temperature 1µK larger (pulses are initially simulated with no noise and no jitter). In addition, noise-and-jitter-free pulses at the calibration energies and the default bias voltage have been also simulated to fit the gain scale curve.

\begin{figure} [ht]
\begin{center}
\includegraphics[height=7.5cm]{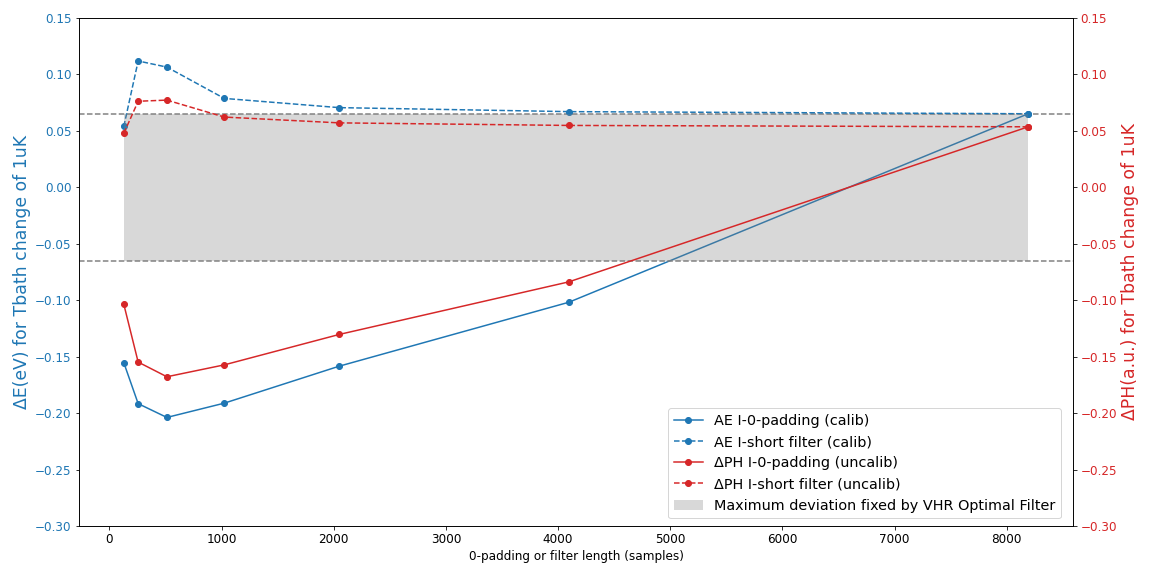}
\end{center}
\caption{0-padding (continuous lines) and short filter (dashed lines) sensitivity to baseline fluctuations. Blue lines represent a change per µK in calibrated energies (left axis, $\Delta$E) and red lines changes per µK in uncalibrated values (right axis, $\Delta$PH). Shaded area dimensions the region for maximum deviation given by the standard (full length) optimal filtering technique.} 
\label{fig:tbath} 
\end{figure}

As for the previous section, both pulses have been reconstructed at different lengths with the 0-padding and the short filters optimal filtering techniques and the results are presented in Fig.~\ref{fig:tbath}. 

In contrast to the previous section, the 0-padding approach would have a sensitivity to changes in the baseline induced by bath temperature changes worse than those shown by the standard (full length 8192 samples) optimal filtering for filter lengths shorter than ~5000 samples.

\subsection{Baseline Level Fluctuations} 
\label{sec:adus}

The sensitivity of the 0-padding technique to baseline level fluctuations can be checked testing its performance over changes of several arbitrary units in this level. As above, a noise-and-jitter-free pulse of 7~keV has been simulated with XIFUSIM, with the default baseline level ($\sim$7000 a.u.). The signal of this pulse has been incremented in 1 a.u. to create the second test pulse. Both pulses have been reconstructed with the standard optimal filter and the 0-padding at different pulse lengths.

\begin{figure} [ht]
\begin{center}
\includegraphics[height=7.5cm]{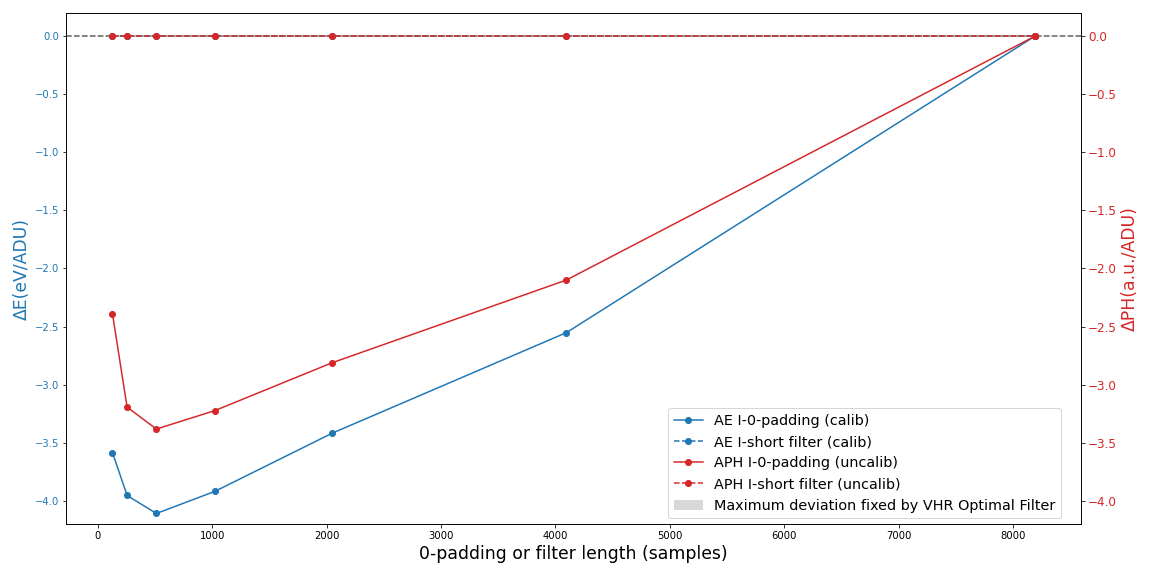}
\end{center}
\caption{0-padding (continuous lines) and short filter (dashed lines) sensitivity to a +1ADU change in the baseline level (no noise data). Blue lines represent a change in calibrated energies (left axis, $\Delta$E) and red lines changes in uncalibrated pulse height values (right axis, $\Delta$PH).} 
\label{fig:adus} 
\end{figure}

Fig.~\ref{fig:adus} shows the magnitude of the change for both reconstructions, being almost 0 for the standard optimal filtering technique (thanks to the filter having 0 sum). However, for the 0-padding reconstruction the change in energy reaches prohibitive values. The $\Delta$E value grows linearly being double after an increment of 2 a.u. in the baseline level.

\section{CONCLUSIONS}
\label{sec:conclusions}

The addition of a pre-buffer signal to the optimal filter and the direct cut of the its length in time domain (0-padding) have been tested as alternative routes to optimize the energy resolution provided by classical short filters built from shorter pulse templates. Despite of the promising results of the 0-padding method in terms of resolution performance with the bonus of diminishing the computational resources needs, its high sensitivity to offset fluctuations (specially to the baseline level) currently prevents its use as the baseline approach for the X-IFU instrument for the lack of consolidated information on the offset stability that will be got in flight. For this reason, classical shorted filters are currently kept as baseline to process pulses below high resolution.

\acknowledgments 
 
This work has been funded by the Spanish Ministry of Science, Innovation and Universities (MICINN) under project RTI2018-096686-B-C21 (co-funded by FEDER funds) and by the Spanish State Research Agency (AEI), Excellency Unit Mar\'{\i}a de Maeztu, MDM-2017-0765. 

\bibliography{report} 
\bibliographystyle{spiebib} 

\end{document}